\documentstyle[11pt,newpasp,twoside,epsf]{article}
\def\mj{M$_{Jupiter}$}

\def\edcomment#1{\iffalse\marginpar{\raggedright\sl#1\/}\else\relax\fi}
\reversemarginpar

\begin{document}
\title{Finding Brown Dwarf Companions with HST/NICMOS}
\author{Patrick Lowrance\footnotemark, Eric E. Becklin}
\affil{UCLA, 405 Hilgard Ave, Los Angeles, CA 90095 USA}
\footnotetext{present address: Infrared Processing and Analysis Center, MS 100-22, Pasadena, CA 91125}
\author{Glenn Schneider}
\affil{Steward Obs., U.Arizona, 933 N.Cherry Ave, Tucson, AZ 86721 USA}
\author{and the NICMOS IDT EONS team \& STIS 8176 team}

\begin{abstract}
We present the results of a HST/NICMOS coronagraphic survey 
for the direct detection of 
substellar companions within the young TW Hydrae and Tucana Associations. 
At the distance of these 
associations, the lower mass limit of detection, based on models, is well into the
high mass planet region for separations $>$ 30 AU . Results presented here include spectra and 
proper motion verification of 
two brown dwarf companions located 100 AU and 180 AU from their primaries.
We also present a possible exo-solar giant planet candidate located 
125 AU from TWA 6. These few examples demonstrate that the 
young associations remain fertile ground for
discovery and environmental study of planetary systems.

\end{abstract}

\section{Introduction}

In the last few decades, a primary goal of observational astronomy has been 
to gain more insight into stellar and planetary formation.  
Brown dwarfs occupy the niche in the mass range
between stars and planets. They form like stars, but do
not have enough mass to sustain hydrogen fusion. The observational distinctions between 
a planet and brown dwarf have yet to be constructed. 
Unfortunately, until just a few years ago, no unambiguous brown dwarfs were known.

Substellar objects cool indefinitely because 
they do not sustain hydrogen fusion and 
thus become fainter and more difficult to detect at older ages (c.f. Burrows et
al. 1997). Therefore, the young 10$-$40 
Myr associations such as TW Hydrae and Tucana represent excellent targets for
a search for brown dwarfs and massive planetary companions.

As part of a larger coronagraphic survey program (Lowrance et al. 2001), 
we surveyed 5 of the members of the TW Hydrae Association,
including TWA 1, 5, 6, 7, 8B, and 10 as well as 2 members of the Tucana Association 
for possible brown dwarf companions, low-mass stellar companions, 
and dust debris disks (see Schneider et al. (2001); this volume). TWA 8A was 
included in another NICMOS imaging survey (Weintraub et al. 2000). 
All other TWA members from Webb et al. (1999) were close multiple systems and we did 
not observe these since other close companions can presumably 
be ruled out on dyanamical grounds.

\section{Using the coronagraph aboard NICMOS}

The main problem with trying to image brown dwarfs or giant planets around
main-sequence stars is the overwhelming brightness of the primary. 
A substellar companion will
be much fainter than the star it orbits (i.e. a Gl
229B-like object, L$=$2$\times10^{-5}$L$_{\odot}$, orbiting a
solar-like star of 1 L$_{\odot}$). The cool 
brown dwarf makes up a little of this in the infrared, where it radiates most
of its power, and is brighter with youth, but the primary is still
much brighter. 

To facilitate the removal of background light from the primary near the coronagraphic 
hole, we designed pairs of observations with different 
spacecraft orientations ($\delta\theta$ = 29.9) which could then be 
subtracted from one another to cancel scattered light. Each observation was 
about 800s long split into 3 Multi-Accum (non-destructive read) sets (see Lowrance et al. 1999).

\begin{figure}
\plottwo{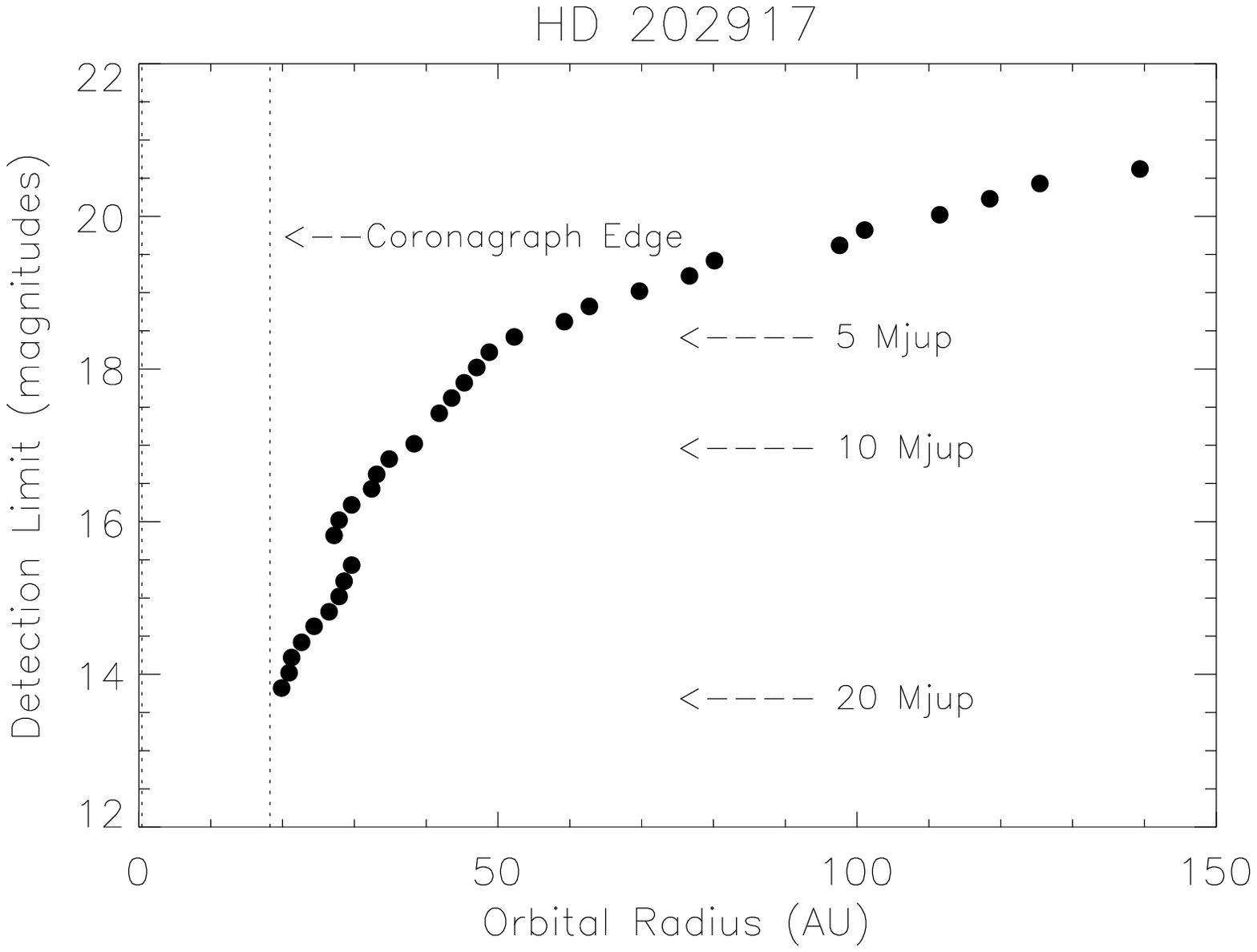}{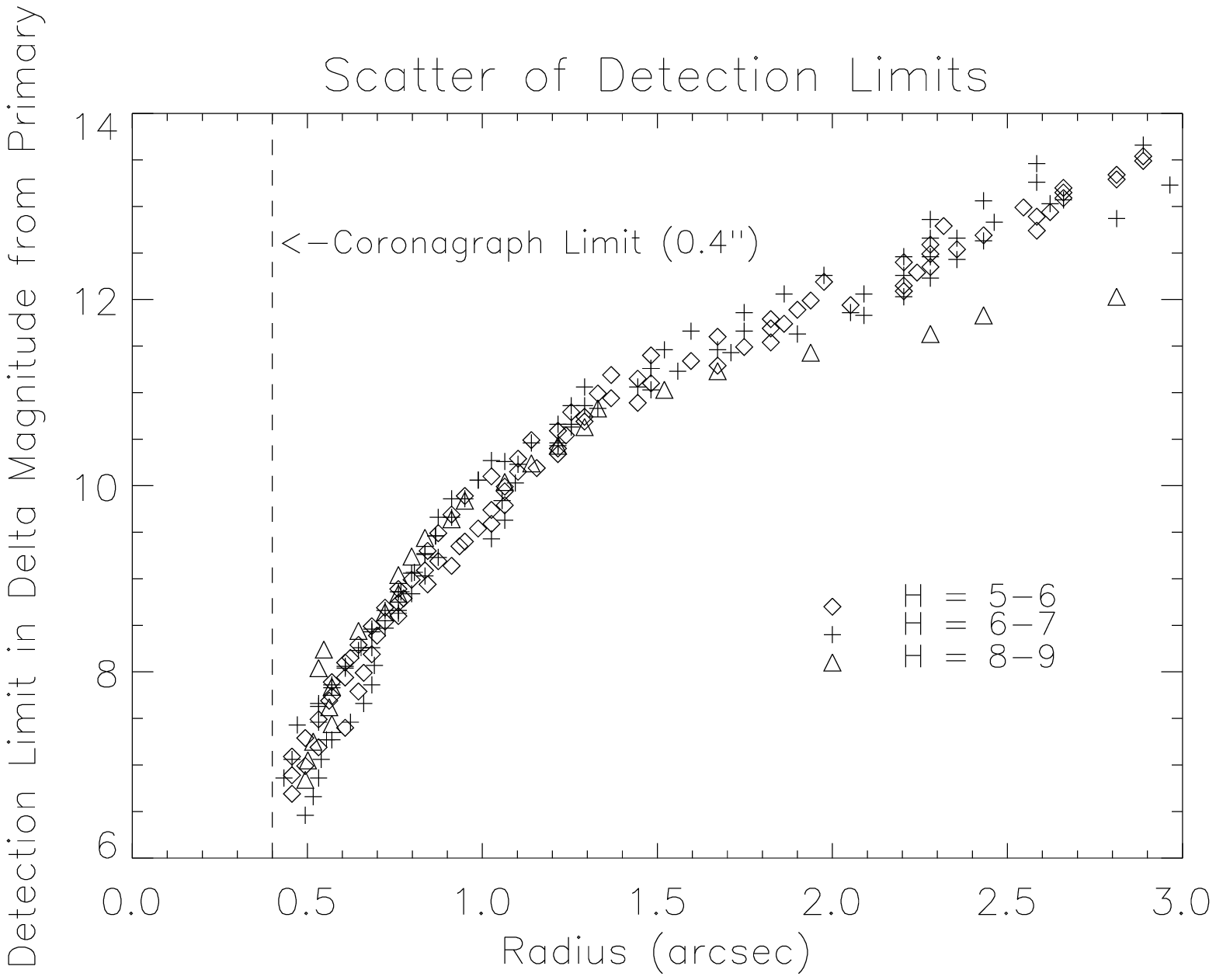}
\caption{Detection limits for point sources around one primary, HD 202917, ({\it left}) 
and overall ({\it right}) in the survey. Masses associated with absolute magnitudes are 
derived from models (Burrows, A. pers comm).}
\end{figure}

In Figure 1, we plot the detection limits for the example of HD 202917 (left) and 
the limits overall (right) in
the observations found from planting and recovering PSF stars in the images. 
At 1$\arcsec$, we can detect a delta magnitude of 9.5 mag for all 
stars. For stars in the Tucana and TW Hya associations, the average primary is 
H $=$ 7 mag, and our average limit corresponds
to M$_H$ = 13.6 mag at 50 AU (median distance is 50 pc). 
As these stars are very young (10--40 Myr), 
the detection limit is $\sim$5\mj, based on the models of Burrows (pers comm). 
Even at 0.5$\arcsec$, we can detect a
delta magnitude of 6--8 mag, a full 2--4 mag better than most speckle
imaging programs. 

In Table 1 we present the results of the coronagraphic
survey. `Stellar-like' candidates are those that have a FWHM between
0.14 and 0.18$\arcsec$, the brighter of which show an Airy diffraction pattern. The
high resolution of the observations 
makes it easy to distinguish between stars and 
diffuse background galaxies. All of the candidate companions 
have been re-observed with adaptive optics to confirm 
companionship and their possible substellar
nature. 

\begin{table}
\caption{Point Sources observed around TWA and Tucana stars}
\vspace{0.1cm}
\begin{tabular}{lccccc}
\tableline
      &Primary &  $<$----- & Secondary  & ----$>$   \\
\tableline
Star  &  Sp Type  &  Sep ($\arcsec$)  & $\Delta$H mag  &  Follow-up  & Results\\
\tableline
TWA 1 &  K7 & none  & --  & --  & --  \\
TWA 5 &	 M3 & 1.96  & 4.9   & AO, STIS & brown dwarf \\   
TWA 6 &  K7 & 2.54  & 13.1   & NICMOS, AO & inconclusive\\
TWA 7 &  M4 & 2.47  & 9.6   & NICMOS, AO & background\\
TWA 8B & M3 & none  & --   & -- & -- \\
TWA 10 & M3 & none  & --   & -- & -- \\
HR 7329 & A0  & 4.17 & 6.9 & AO, STIS & brown dwarf\\
HD 202917 & G5  &  none  & --  & -- & -- \\
\tableline
\tableline
\end{tabular}
\end{table}

\section{Brown Dwarf Companions}

\begin{figure}
\plottwo{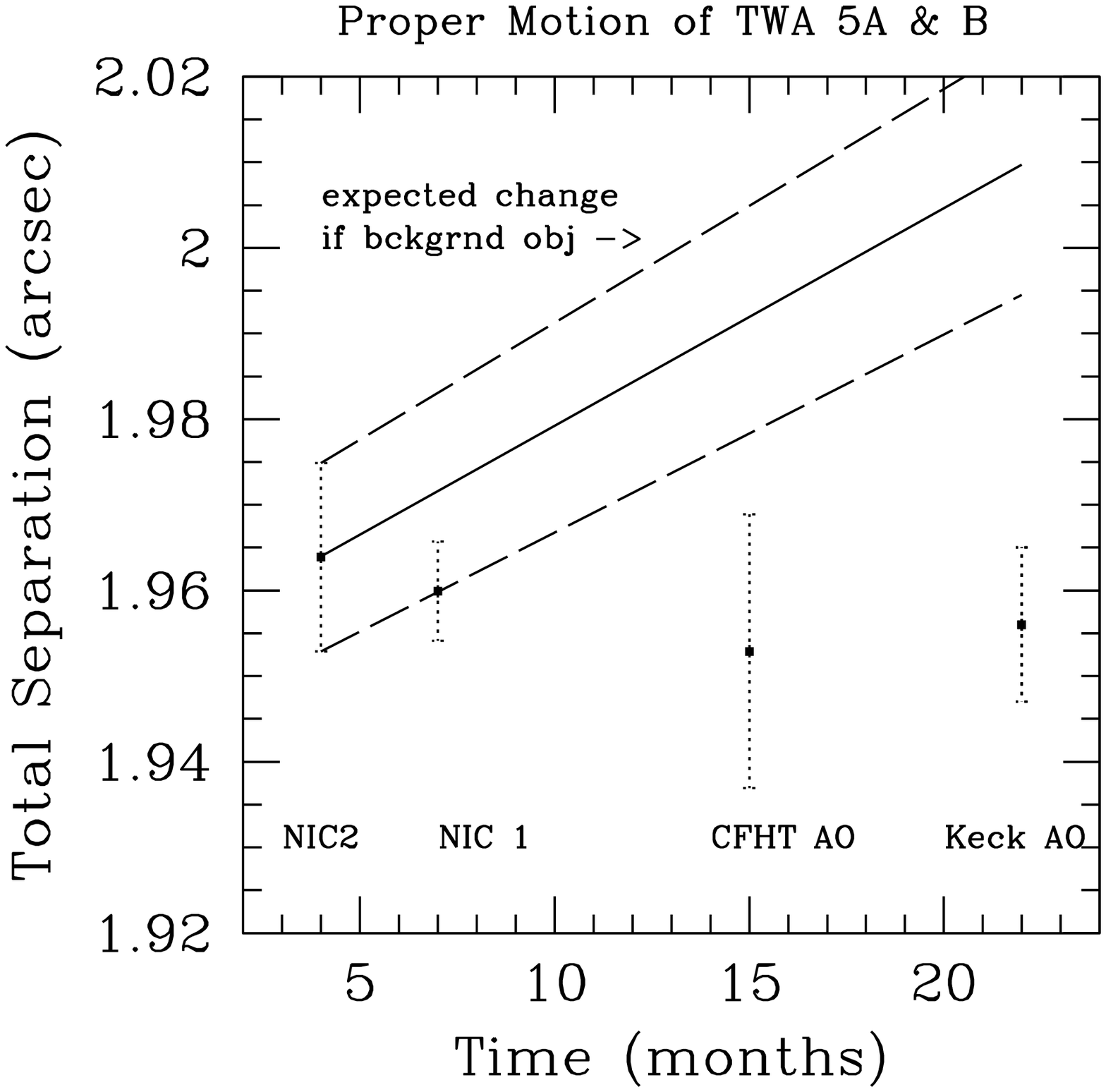}{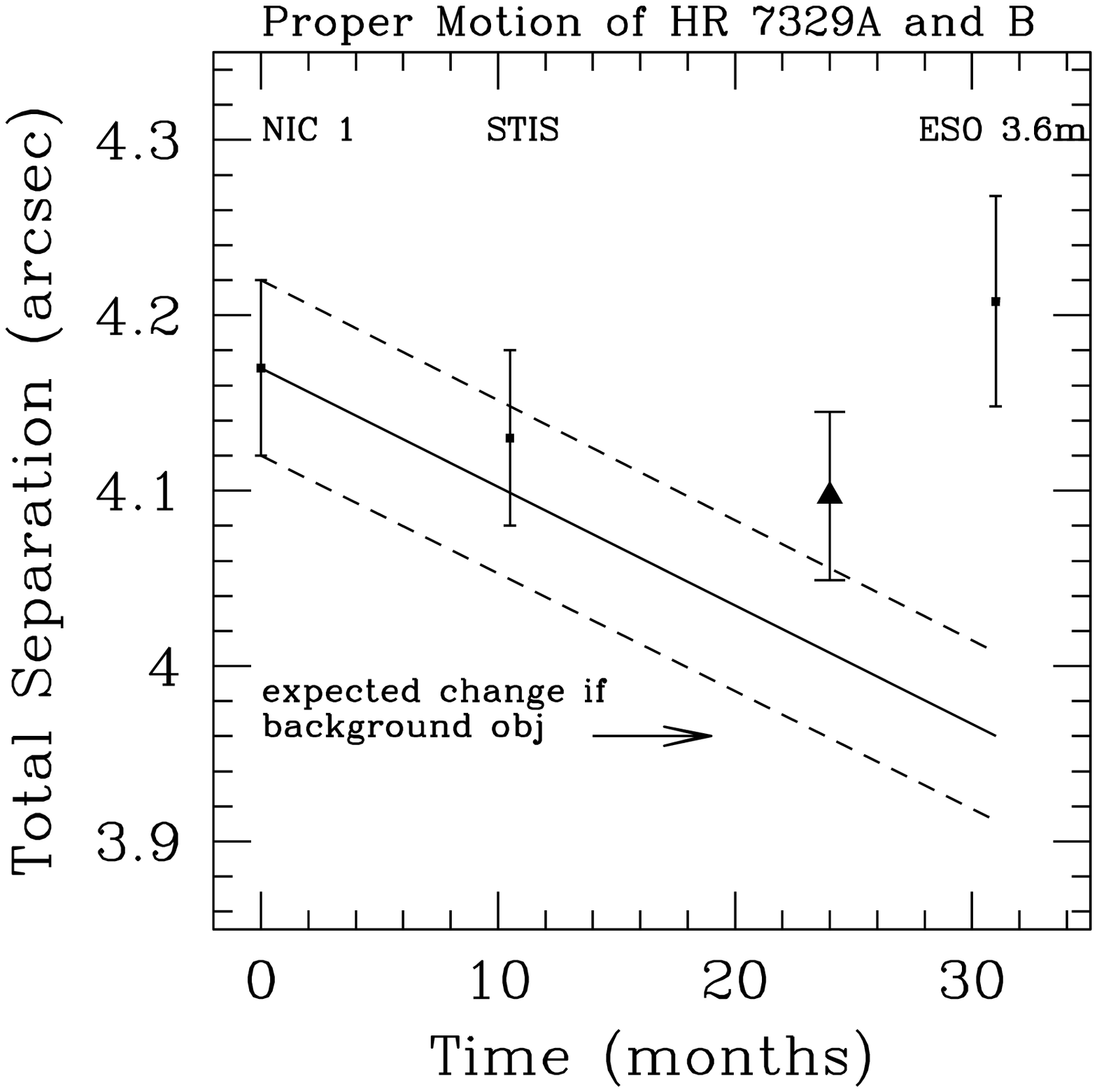}
\caption{Measured separations of TWA 5A \& B ({\it left}) and 
HR 7329A \& B ({\it right}) over the last two and a half years. 
The dashed lines represent the change in separation expected if the candidate brown dwarf, B, 
is a stationary object in the background based on the measured proper motion (and
error) of the primary.} 
\end{figure}

\subsection{Astrometry}

The TWA 5 system was observed 
several times in the last two years -- with NIC 1 (Weintraub et al. 2000) 
and with the AO systems on the Canada-France-Hawaii telescope (CFHT) and the 
10-m W.M. Keck II telescope (Figure 3(a)). The HR 7329 system was observed 
with the Space Telescope Imaging Spectrograph (STIS) instrument and using the 
AO system on the 3.6m ESO 
telescope. We also include the measurement of HR 7329 
from Guenther et al. (2000) as the triangle in Figure 3(b). 
From these observations, we plot the 
change in measured separation of the A and B components at each epoch 
and compare them with the expected change in separation for a 
background object. We conclude both TWA 5B and HR 7329B 
are true proper motion companions to their primaries. 

\begin{figure}
\plottwo{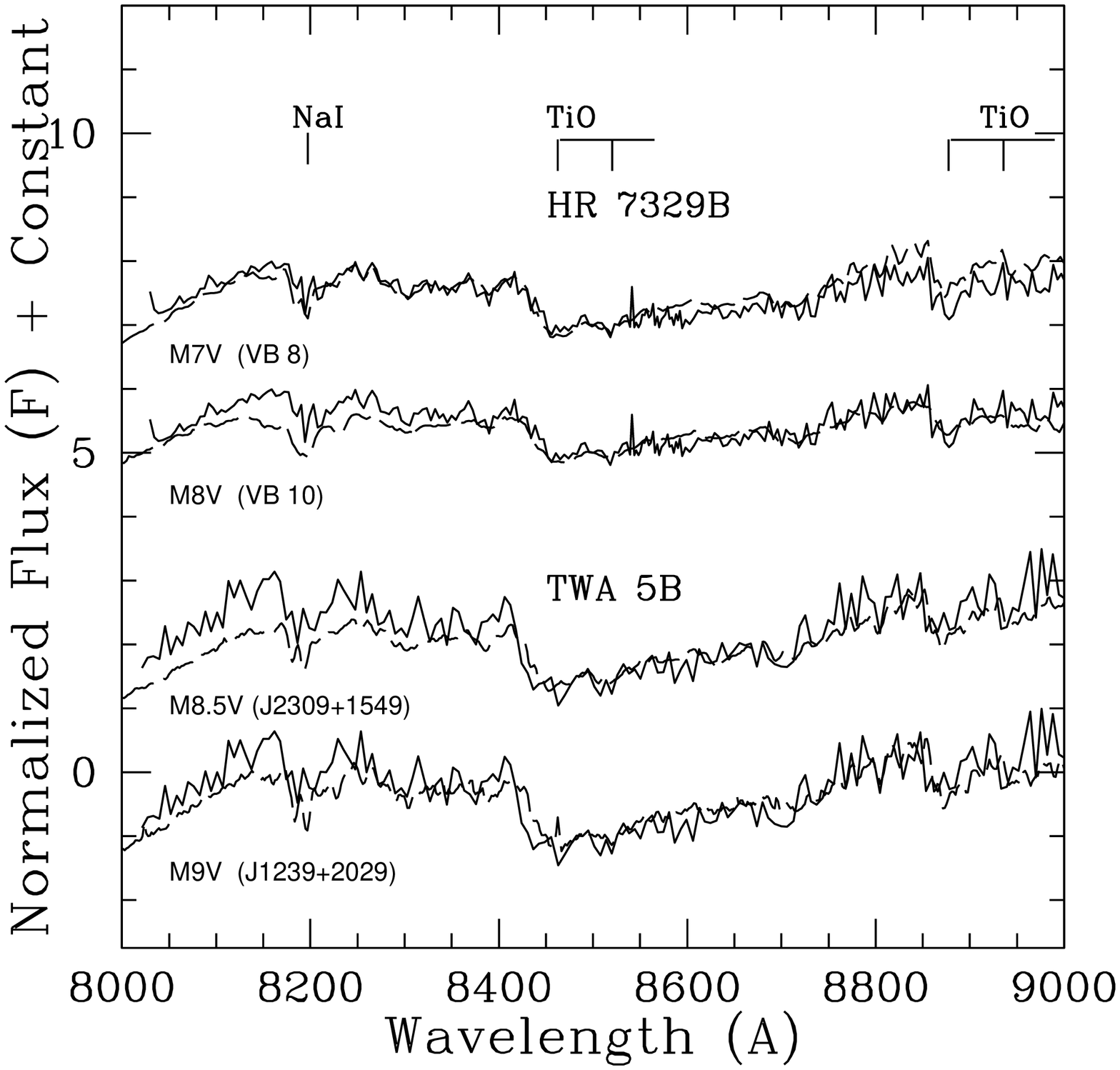}{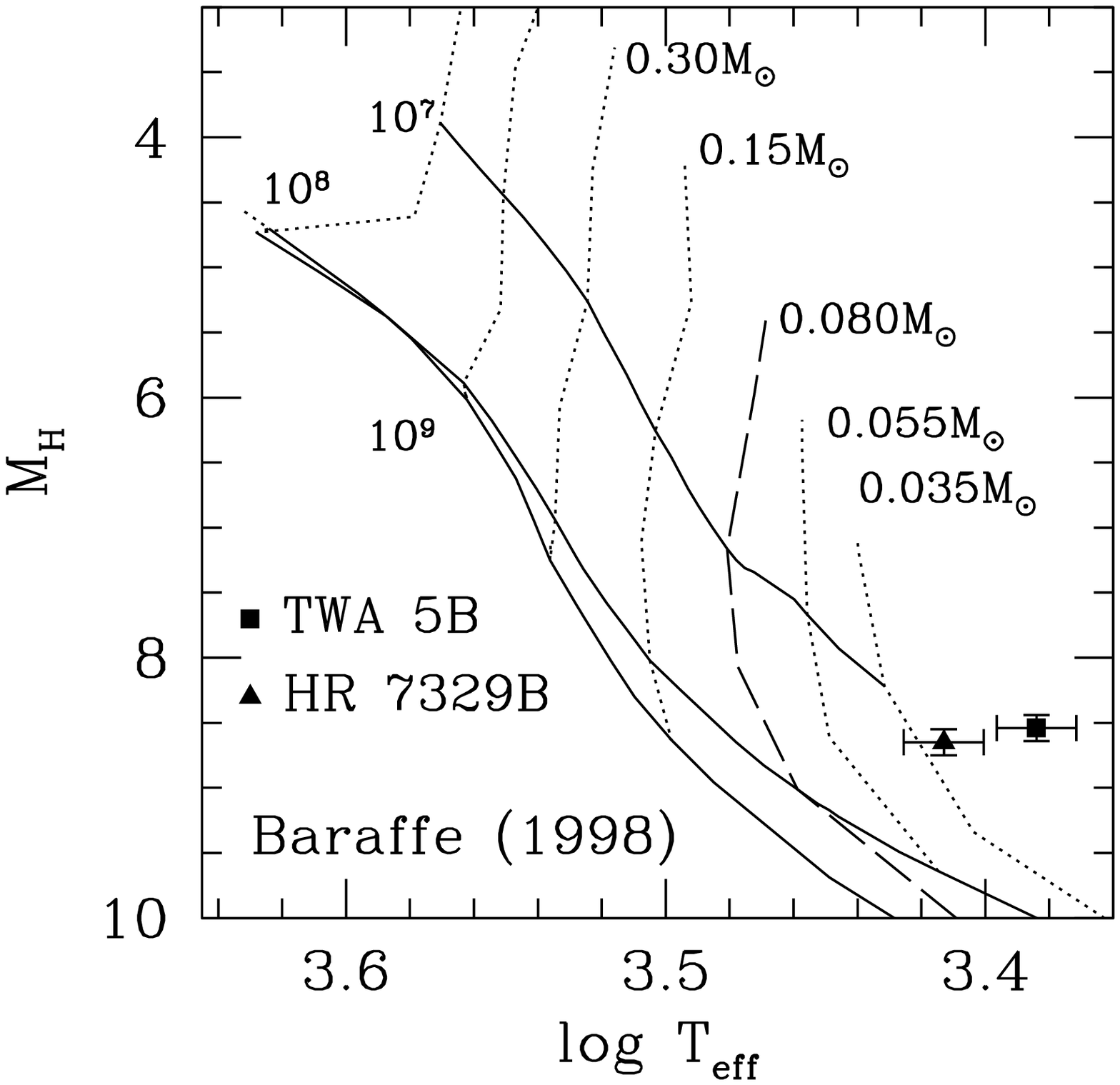}
\caption{STIS spectra({\it left}) of TWA 5B and HR 7329B compared with template 
late M-dwarf spectra. ({\it right}) The location of these two objects on 
theoretical evolutionary diagrams.}
\end{figure}

\subsection{STIS spectra}

The companions to TWA 5A and HR 7329B were placed in the slit of 
the STIS 
by centering the primary and offsetting to the companion by the NICMOS-measured astrometry 
(Lowrance et al. 1999, Lowrance et al. 2000). 
Spectral imaging sequences were
completed in one orbit with the G750M grating. 
The STIS spectra of TWA 5B and HR 7329B were fit to template M dwarf spectra 
(Figure 2), and classified as 
M9 $\pm$ 0.5 for TWA 5B and M7.5 $\pm$ 0.5 for HR 7329B.  For TWA 5B, this is consistent with the
photometric spectral type of M8-M8.5V derived by Lowrance et al. (1999) 
and more recent results by Neuhauser et al. (2000).

Using the temperatures derived from the spectral types and absolute H magnitudes 
which assume the same distance as the primary, we place these two objects on 
evolutionary diagrams of Baraffe et al. (1998)(Figure 2(b)). 
TWA 5B is consistent with a 20 \mj\ object at 
an age of 10 Myr old, which is the approximate age of the TW Hydra Association. HR 7329B 
is consistent with a 
40\mj\ object at an age of 30-40 Myr, the approximate age of the Tucana Association. 

\section{TWA 6B - giant planet candidate}

A point-source was discovered at 
a separation of 2.549$\arcsec$ $\pm$ 0.011, and a position angle of 
-278.7$^{\circ}$ $\pm$ 0.2 
from TWA 6 (TWA 6A). The H magnitude of TWA 6B is 19.93 
$\pm$ 0.08 mag. The field of TWA 6 was reobserved with the NICMOS 1 camera with a
medium-band F090M filter (central wavelength: 0.9003~$\mu$m,
$\Delta\lambda$ = 0.1885~$\mu$m) and TWA 6B was not detected. We derive an upper
limit (3$\sigma$) to the flux of [F090M]=22.6 mag in the predicted
position from the NICMOS images. 
Using low-temperature models to transform between F090M and I-band, we 
calculate an upper limit of I$-$H$>$3.3 for the candidate companion. 
The color is equivalent to a spectral type later than 
M7V (Kirkpatrick \& McCarthy 1994), so we conclude that the
object is very red, even if it is not associated with TWA 6A. 
A background K giant would have an I$-$H $<$ 2 mag, which 
would have been easily detected in the NIC 1 images. If associated with
TWA 6A at 50 pc, this object would have an absolute magnitude,
M$_H$=15.8, which corresponds to a $\sim$ 2\mj\ object at 125AU
(Burrows, A. pers comm). 

Re-observations of TWA 6 are underway with the Keck 
AO system (Macintosh et al, this volume). Such 
astrometric measures are difficult with the current 
two year baseline, but high resolution observations 
are needed to establish if TWA 6B 
is a young Jovian planet.

\section{Conclusions}

With the NICMOS camera and the coronagraph, we have studied the environments 
of close to half of the young
stars of the TW Hydrae Association and two of the Tucana Association. 
Around four of these stars we found
point-like objects which were possible companions. After careful
analysis and more observations, we find that TWA 5B is a $\sim$ 20
\mj\ brown dwarf, HR 7329B is a $\sim$ 40\mj\ brown dwarf  
and TWA 7B is a background object. We find no
candidate companions around HD 202917, TWA 1, TWA 8B, or TWA 10 greater than a
few \mj, based on models of giant planets, at 1$\arcsec$, or 50 AU at
the distance of the stars. A faint, point-like object
2.5$\arcsec$ from TWA 6 remains a mystery, though. 
It has the potential to be the most significant discovery of this
program as the first giant planet imaged around another star.

\end{document}